\documentclass[10pt,oneside,reqno]{amsart}

\usepackage{amstext,amsmath,amsfonts,amscd,amsthm,graphicx}

\theoremstyle{plain}

\newtheorem{definition}{Definition}

\theoremstyle{remark}
\newtheorem{remark}{Remark}

\begin{document}
\title[Spectral Quadruples]{Spectral Quadruples${ }^{1}$}\thanks{${ }^{1}$
Talk presented at the Euroconference
      "BRANE NEW WORLD and Noncommutative Geometry",
          Torino, Villa Gualino (Italy), October 2 - 7, 2000}
\author{Tom\'{a}\v{s} Kopf}
\address{Matematical Institute of the Silesian University at Opava,
Bezru\v{c}ovo n\'{a}m\v{e}st\'{i} 13, 746 01 Opava, Czech Republic}
\curraddr{ThEP, Institut f\"{u}r Physik, Johannes
Gutenberg-Universit\"{a}t,
55099 Mainz, Germany}
\email{kopf@ThEP.Physik.Uni-Mainz.DE}
\thanks{First author supported by the Alexander von Humboldt Foundation}

\author{Mario Paschke}
\address{ThEP, Institut f\"{u}r Physik, Johannes
Gutenberg-Universit\"{a}t,
55099 Mainz, Germany}
\email{paschke@ThEP.Physik.Uni-Mainz.DE}





\begin{abstract}
A set of data supposed to give possible axioms for spacetimes.
It is hoped that
such a proposal can serve to become a testing ground on the way to
a general formulation.
At the moment, the axioms are known to be sufficient for cases with a 
sufficient number of symmetries, in particular for 1+1 de
Sitter spacetime.   
\end{abstract}

\maketitle


A major shortcoming of applications of noncommutative geometry to high energy physics (including gravity) so far has been the Euclidean formulation of such models. E.g., the fermion doubling in the noncomutative description of the standard model can be traced to the use of a metric with Euclidean signature and concepts like {\em causality} are needed for the formulation of realistic  quantum field theories. 

Another important problem, closely connected with the one above is the absence of an action principle. One may formulate an "action" through a Lagrange density integrated over the entire (Euclidean) space \cite{Connes96a}. For the moment that space can only be compact and boundary values cannot be incorporated. The solutions of the equations of motion  (respectively the extramal points of the action) are then naturally non-unique.

For a complete action principle, one should be able to give the values of the fields at on two arbitrary spacelike hypersurfaces, i.e.  the initial and final field configuration,  as boundary conditions under which the extremum of the action is to be found. This goal is at the moment still quite far away.

Globally hyperbolic spacetimes of dimension $d+1$ (physically, the Lorentz signiture is relevant only) may always be written as $\Sigma\times \mathbb{R}$, where $\Sigma$ is a $d$-dimensional manifold. This corresponds to a foliation of the spacetime along the time axis $\mathbb{R}$. The spacelike hypersurface ${\Sigma}_{t}$ is at each time homeomorphic to $\Sigma$.

It is then at hand to deal with both of the above sketched problems by working with a $(d+1)$-splitting of spacetime and to describe the spacelike hypersurfaces ${\Sigma}_{t}$ by spectral triples. The time coordinate $t$ is then understood as a parameter and thus a whole family of spectral triples is obtained. However, attempts to transfor these ideas into an axiomatics for "noncommutative causal (Lorentzian) spin manifolds" have up to now had little success. One reason for this is certainly the technical difficulty of translating particular toy models into an algebraic language. As in the Euclidean case,  also here the additional symmetries of a homogeneous (with respect ot the symmetry group)  spacetime turn out to be precious tool. Above that, many physically important examples like the Robertson-Walker solutions of the Einstein equations have a high degree of symmetry.

In \cite{Kopf-Paschke2000b}, we described algebraically and completely such toy models. This had led to a working hypothesis on the axioms for a causal spin manifold which we called a spectral quadruple. These axioms are motivated, sketched and discussed below.

As a starting point, the question of the correct Hilbert space for the spectral triple of the hypersurface ${\Sigma}_{t}$ or better of the representation ${\pi}_{t}$ of the algebra $C({\Sigma}_{t})$  will be clarified. There is nothing to say against the choice $\mathcal{H}_{t}={L}^{2}({\Sigma}_{t}, {S}_{t})$ for each time $t$ where ${S}_{t}$ is the restriction of the $d+1$-dimensional spinor bundle over ${\Sigma}_{t}$. This would lead to a family of Hilbert spaces which would then have to be "glued together"  in a suitable way to be able to describe a smooth time evolution.  However, a simple and elegant way to implement time evolution is offered by the Hamiltonian of the spinor field:

In the $d+1$-formalism, the Dirac equation on the entire spacetime can written  in Hamiltonian form,
\begin{align}
H\psi &= i{\partial}_{t} \psi
\end{align}  
where the Hamiltonian $H$ is given at each time by
\begin{align}
H&=-iN\left( {\omega}_{0}^{s}+ {\gamma}_{a} {\gamma}_{b}{e}_{0}^{a}{e}_{i}^{b}{g}^{ij} {D}_{j}^{s} - m{\gamma}_{a}{e}_{0}^{a}    \right) +{N}^{i}{\partial}_{i} .\label{hamiltonian}
\end{align}
Here, ${\omega}_{\mu}^{s}$ denotes the components of the spin connection, ${D}_{\mu}^{s}$ the corresponding covariant derivative and ${\gamma}_{a}$ the generators of the Clifford algebra $\{ {\gamma}_{a} ,{\gamma}_{b} \}=2{\eta}_{ab}$.

${e}_{\mu}^{a}$ is the local $(d+1)$-bein and is chosen in such a way that its spatial components ${e}_{j}^{a}$  are at each time $t$ tangential to ${\Sigma}_{t}$ while ${e}_{0}^{a}$ is orthogonal to ${\Sigma}_{t}$. The tangent vector $\frac {\partial}{\partial t}$ of the (arbitrarily chosen) time axis can be decomposed as 
\begin{align}
{\partial}_{t}&=N{\partial}_{0} + {N}^{i}{\partial}_{i}
\end{align} 
which at the same time defines the lapse function $N$ and the shift vector ${N}^{i}$. These describe the $t$-direction (relative to the given hypersurface ${\Sigma}_{t}$ and to the orthogonal $0$-direction) and express the arbitrariness in its choice. (The vector ${N}^{i}$ basically reflects the free choice of coordinates on ${\Sigma}_{t}$.)

Therefore, there are clearly  different Hamiltonians which describe the same spacetime. This has to be taken into consideration in the following.

Using $H$ one can now define the time evolution operators 
\begin{align}
{U}_{H}({t}_{1},{t}_{2})&: {\mathcal{H}_{t_1}}={L}^{2}({\Sigma}_{t_1}, {S}_{t_1})\rightarrow
{\mathcal{H}_{t_2}}={L}^{2}({\Sigma}_{t_2}, {S}_{t_2})
\end{align}   
which can be used to identify the different Hilbert spaces \cite{Hawkins97}. Corresponding elements ${\psi}_{t_1}$, ${\psi}_{t_2}$ are identified if and only if they are restrictions of the same solution of the Dirac equation to ${\Sigma}_{t_1}$ respectively ${\Sigma}_{t_2}$.

Naturally, it is more elegant and easy to work on a single Hilbert space , the space of solutions of the Dirac equation, i.e., the phase space of the spinor field. This space is isomorph to ${L}^{2}({\Sigma}_{t}, {S}_{t})$, since a solution is uniquely determined through its values on a spacelike Cauchy surface. At the same time, it can be understood as the one-particle subspace of the Fock space of the quantum field theory of the spinor field. This is a very desirable side effect if one considers that one of the great goals of noncommutative geometry is to derive the geometry of spacetime from a (complete) quantum theory.
The spectral quadruples introduced in the following are thus also a first step in a zeroth approximation by describing classical spacetime in a language adjusted to quantum field theory. Hereby, spacetime is, however, still understood as a fixed background rather than as a dynamical variable which ought to be quantized.

Given on this Hilbert space $\mathcal{H}\cong {L}^{2}(\Sigma , S)$ (the spin bundles are at all times equivalent) at an arbitrary time  $t_0$ the remaining data $({D}_{t_0}, {\pi}_{t_0}(C(\Sigma )), {\gamma}_{t_0}, J)$ of a spectral triple then one can construct the spectral triples at all other times $t$ through the time evolution operators ${U}_{H}( t_1 , t_2 )$ by transporting any of the involved operators $\mathcal{O}_{t_0}$
\begin{align}
\mathcal{O}_{t} &= {U}_{H}( t_0 , t )\mathcal{O}_{t_0} {U}_{H}( t , t_0 )
\end{align}    
from $t_0$ to $t$. Of course, this presupposes that the time evolution operator ${U}_{H}( t_1 , t_2 )$  and thus the Hamiltonian $H$ are explicitly given. But then a single look at (\ref{hamiltonian}) is sufficient to realize that it should be possible to derive the spatial Dirac operator
\begin{align}
{D}_{t} &= {\gamma}_{a} {e}_{i}^{a} {g}^{ij} {D}_{j}^{s}
\end{align}
on the hypersurface $\Sigma_t$. It will therefore not appear in the data of the spectral quadruple. How it may be recovered from these data will be discussed later.

Working in the $(d+1)$-formalism, in particular with the Hamiltonian, there is always an arbitrariness of the time direction to be considered. It has to be clarified how spectral quadruples describing the same spacetime are to be identified. For spectral triples, this happens naturally through the notion of unitary equivalence. Diffeomorphisms are in this case represented by unitary operators on the Hilbert space. But not all diffeomorphisms of the spacetime $\Sigma\times\mathbb{R}$ are unitarily represented on the Hilbert space ${L}^{2}(\Sigma , S)$. (This is from the algebraic point of view the reason for the appearance of $N$ and $N^i$. ) Till now we could not find a solution of the problem to identify equivalent Hamiltonians.

This can though be circumvented in a simple way, paradoxically extending the data of the quadruple.Taking instead of one Hamiltonian which would be entirely sufficient to reconstruct spacetime {\em all} Hamiltonians, on gets in a trivial way a unique description of  spacetime, a rather redundant one. In this way, the covariance of the formulation under diffeomorphisms is ensured.

This may look like a card trick but one should not forget that the (finally) following axioms are to be understood as a working hypothesis only. They are not a final formulation. The advantage of these axioms compared to earlier works is that they allow the explicit completion of particular examples like the de Sitter spacetime. With the help of these examples, a solution to the above mentioned problem of covariance should be found to eradicate the present redundance of the formulation. (To this end, one would in corresponding examples anyways need different, equivalent Hamiltonians on the same Hilbert space.)

Before giving the axioms of a spectral quadruple, a notional explanation is necessary to allow a short and elegant formulation.
Given a category (a collection of objects and morphisms between objects that can be composed, satisfying a certain set of
axioms), a subset of morphisms such that each of its elements has an inverse morphism forms a groupoid.
The algebras $\mathcal{A}_{t}=C({\Sigma}_{t})\cong {\pi}_{t}(C({\Sigma}_{t}))$ can be considered as objects of a category with unitary equivalences between the algebras as morphisms. In particular, the time evolution operators for each (generally time-dependent) Hamiltonian form a subset of these morphisms and since their inverses also exist they form with them a grupoid.

An operator needed in the data of the spectral quadruple and not mentioned yet is the time-direction operator $E$, identified in the commutative case with ${\gamma}_{a} {e}_{0}^{a}$. With its help, the correct commutation relations of the Clifford algebra ${Cl}_{d,1}$ will be obtained and the time orientability of the described manifold will be ensured.  Strictly speaking, only its existence  has to be required, since it then can be recovered from the remaining data. But it is simpler to include it into the already too mighty  flood of data. It is the presence of this operator that makes out of a triple a quadruple.

Since the definition of the spectral quadruple does not concern a single time (a single Hamiltonian)  only, it is better to write, e.g., $\mathcal{A}_{\bullet}$ instead of $\mathcal{A}_{t}$. The index $\bullet$ labels elements of an index set that contains all allowed time evolutions.

\begin{definition}(\cite{Kopf-Paschke2000b})\label{defspectquad}
A {\bf spectral quadruple} $({A}_{\bullet}, \mathcal{H}, G, C, \gamma (\bullet) ,{E}(\bullet))$ consists of a collection of
algebras ${A}_{\bullet}$
represented on the Hilbert space $\mathcal{H}$, of a groupoid $G$ and of an antilinear operator $C$.  In addition, for each
of the algebras ${A}_{\bullet}$ two operators $E$, $\gamma$ are given. These structures
satisfy the following conditions:
\begin{enumerate}
\item {\bf Evolution}.

Any two algebras ${A}_{0}$, ${A}_{1}$ of the collection ${A}_{\bullet}$
are required to be mutually unitarily equivalent through a (not
necessarily unique) unitary $U({A}_{0}, {A}_{1})$ and {\em not} mutually commutative,
$[{A}_{{t}_{1}},{A}_{{t}_{2}\neq {t}_{1}}]\neq 0$.
The groupoid $G$ consists of a subset of all
possible unitary equivalences between the algebras in the collection ${A}_{\bullet}$. It is assumed that for
each algebra ${A}_{0}$ in the collection there exists an evolution, a (not necessarily unique) differentiable
path ${\alpha}_{t}:t\in\mathbb{R}\rightarrow {U}_{t}({A}_{0}, {A}_{t})$
with ${\alpha}_{0}= 1$ such that the generator (derivative) at $t=0$, denoted by
$iH$ is compatible with the further requirements.

\item {\bf Charge conjugation.} The antilinear operator $C$ commutes with $G$ and satisfies
\begin{align}
{C}^{2} &= {(-1)}^{s(n)}\\
\intertext{for the spacetime dimension $n$ and with}
s(n)&:=\frac {1} {8} (n-1)(n-2)(n-3)(n-4). 
\end{align}

\item {\bf First order condition (Dynamics)}.
\begin{align}
\left[ \left[ f, iH\right] , {g}^{op}\right] &=0& \text{for any $f,g\in{A}_{0}$ and any generator $iH$,}
\end{align}
with ${g}^{op}=C{g}^{\ast}C$.

\item {\bf The time vector}. For each algebra ${A}_{0}$ in the collection ${A}_{\bullet}$ there exists an operator $E$ called the time vector satisfying
\begin{align}
{E}^{2} &= -1\\
{E}^{\ast}&=-{E}
\end{align}
and the compatibility conditions in 5. and 6. of this definition.

\item {\bf The volume element}. For any ${A}_{t}$ there exists an operator $\gamma$ such that
\begin{align}
{\gamma}^{2}&=\pm 1&\\
\{ {E},\gamma \} &= 0 &\text{for even spacetime dimension}\\
[ {E},\gamma ] &= 0 &\text{for odd spacetime
dimension}
\end{align}
and
\begin{align}
 \gamma &= {E}
\sum_{{f}_{\bullet}\in {A}_{t}}{{f}_{i_0}[\tilde{D},{f}_{i_1}]\ldots
[\tilde{D},{f}_{i_n}]}
&\text{for even spacetime dimension $n+1$}\label{hoch1}\\
\gamma &= \sum_{{f}_{\bullet}\in {A}_{t}}{{f}_{i_0}[\tilde{D},{f}_{i_1}]\ldots [\tilde{D},{f}_{i_n}]} \label{hoch2}
&\text{for odd spacetime dimension $n+1$}
\end{align}
\begin{align}
\intertext{and for suitable functions ${f}_{\bullet}$ where $\tilde{D}$ is given by}
\tilde{D}&=
\begin{cases}
\gamma [iH,\gamma ]&\text{for even spacetime dimension}\\
iH&\text{for odd spacetime dimension}\\
\end{cases}&
\end{align}

\item {\bf Geometry of space.} For any algebra ${A}_{t}$ of the collection
$({A}_{t}, \mathcal{H} , D=E [H, E], \gamma , C)$ is
\begin{itemize}
\item a spectral triple for odd spacetime dimension.

\item a spectral triple for even spacetime dimension, if restricted to each of the two eigenspaces of $E$.
\end{itemize}
\end{enumerate}
\end{definition}

\begin{remark}
The operators $\tilde{D}$ used in condition 5. are in no way identical to the Dirac operator. (That comes from the term $N^i {\partial}_i$ in $H$.) However, for $n$ odd, these terms disappear from the sum defining the Hochschild cycle due to antisymmetrization. For $n$ even, $\tilde{D}$ has the same principal symbol and is therefore in the definition of the Hochschild cycle equivalent to the Dirac operator. The same applies to the operator $D=E [H, E]$ which is used in the definition of the spectral triple on a hypersurface. It differs from the "true" Dirac operator by the addition of a term of order zero only which does not play a role in the axioms of spectral triples. But because of 5., it satisfies automatically
\begin{align}
D\gamma &= (-1)^n \gamma D
\end{align} 
\end{remark}

\begin{remark}
Of course, it is allowed to choose $G\cong \mathbb{R}$. Then there would be only one (possibly time dependent) Hamiltonian and only one time direction. In this case it would also suffice to give as data for the spectral quadruple the algebra at time $t=0$ (and correspondingly ${\gamma}_{0}$ and $E_0$) only. A minimal (real, even) spectral quadruple would then be given by $(\mathcal{H}, {\mathcal{A}}_{0}, iH, {E}_{0}, {\gamma}_{0})$ and the difference, compared to a spectral triple, would be "only"  the addition of the operator $E$ (and of course the replacement of the Dirac operator on the hypersurface by the Hamiltonian. Given $E$, this does not make much of a difference.)

However, then one would have to worry about the missing covariance of the formulation again. Moreover, new possibilities to transceed the situation of a foliated manifold follow from the choice of a larger $G$.

There is, however, a problem with a large choice of $G$, in the extreme case all possible time evolutions. The larger $G$ is, the more algebras $\mathcal{A}_{t}$ are needed in the data of the spectral quadruple. These algebras correspond to functions on the hypersurfaces ${\Sigma}_{t}$ which arise by time evolution from ${\Sigma}_{0}$ (corresponding to ${\mathcal{A}}_{0}$). If there is just one Hamiltonian then all appearing hypersurfaces may be disjunct. They can be constructed as the spectrum of the algebras ${\mathcal{A}}_{t}$ (when these are commutative) and can be glued together along the time axis. But when there are more equivalent time evolutions this is no more the case. The corresponding hypersurfaces will mutually intersect and this must be properly considered in a reconstruction of spacetime. To compare characters (i.e., points) of commutative ${C}^{\ast}$-algebras, one has to extend these characters to a larger algebra that contains all relevant algebras as subalgebras. But this is generally not possible, if the set of algebras is chosen too large. Thus, the set of algebras and together with it $G$ has to be restricted (possibly by a smoothness principle.)   
\end{remark}

\begin{remark}
It is not obvious that all Hamiltonians derivable from $G$ describe the same manifold. It appears to be allowed to use in one spectral quadruple different Hamiltonian operators corresponding to different metrics. This is however not the case due to the grupoid structure of $G$. It has to be always possible to compose unitary equivalences $U(A,B)$ and $U(C,D)$ if the algebras $B$ and $C$ are identical. (There is, of course, the possibility to have two non-equivalent Hamiltonians in $G$ in the sense that the spectral quadruple can be decomposed into the direct sum of two spectral quadruples for the corresponding spacetimes. This case can be easily excluded by requiring the irreducibility of the representation of the data of the quadruple, which is naturally employed in the case of spectral triples.)
\end{remark}

\begin{remark}
Even when it is possible to reconstruct spacetime out of the data of the spectral quadruple, there remains the question of whether it is possible to reconstruct also the metric. The total Dirac operator on spacetime is not available, respectively is equal to $m \mathbf{1}$ 

(Even if it where known, the problem would not be an easy one. Connes' distance formula works due to the involved supremum in the Euclidean signature only.) 

But it is anyways in the spirit of the spectral quadruple to remind oneself of quantum field theory in order to solve this problem. 

A pure state on the algebra $\mathcal{A}_0$ (a point in ${\Sigma}_0$) is described by a $\delta$-distribution. By time evolution, the wave function is smeared into the entire light cone.  (If $m\neq 0$. For massless particles it is smeared onto the edge of the light cone only.) Thus it is no more a pure state at a later time $t$, i.e., no eigenvector of the algebra ${\mathcal{A}}_{t}=C ({\Sigma}_{t})$. Since the algebras ${\mathcal{A}}_{t}$, ${\mathcal{A}}_{0}$ have different eigenvectors, they do not mutually commute. This is the reason behind the corresponding requirement in 1. which ensures that there is an observable time at all. (However, for massless particles in 1+1 dimensions, with a one-dimensional boundary of the light cone these stay eigenvectors for all times because sufficient space for smearing is not available. But then one has to ask oneself how one could possibly measure distances in a (1+1)-dimensional world of massless particles only.)

From the commutators of functions at different times, the entire light cone and the metric can be reconstructed. This is easily seen by expanding the commutator of $f_0\in C({\Sigma}_{t_0})$ and $g_1\in C({\Sigma}_{t_1})$ in powers of $(t_0-t_1)$:
\begin{align}
[f_0,g_1] =
                        & -{({t}_{1}-{t}_{0})}^{2}({N}^{2} {g}^{ij}{g}^{kl}({\partial}_{i}f_0){\Omega}_{jk}{\partial}_{l}g)\\
                        & +{({t}_{1}-{t}_{0})}^{3}4Em{N}^{3}{g}^{ij}
                        ({\partial}_{i}f_0)({\partial}_{j}g_1)+
                        \begin{pmatrix} \text{terms vanishing}\\ \text{in $1+1$ dimensions}\end{pmatrix}\label{thirdorder}\\
                        & + O\left({({t}_{1}-{t}_{0})}^{4}\right)
\end{align}
The functions are evaluated on the hypersurface ${\Sigma}_{t_0}$ ($g_1$ must be first transported there to do this) and ${\Omega}_{jk}$ is the representation of the corresponding generator of rotations on spinors. Considering the expectation value of this operator at suitably chosen states, one can (at least in low dimensions) determine the spatial metric $g^{ij}$ on ${\Sigma}_{t_0}$ and the lapse function from the different orders. For the full reconstruction of the spacetime metric, the shift vector is still needed. That can be obtained from
\begin{align}
i[f_0, H] &= NE {e}_{i}^{a} {\gamma}_{a} ({\partial}^{i}f_0) -N^i {\partial}_{i} f_0
\end{align}    

The commutator of functions at different times is at least of second order in $(t_0-t_1)$ because the Hamiltonian is a differential operator of first order. Remarkably, the second order is then independent of the mass and thus of the length scale. From this order, one can thus only infere the conformal structure of spacetime which can be also seen from the appearance of the generators of rotations.
\end{remark}

If one wishes to describe homogeneous spaces like de Sitter space, i.e., spaces on which a Lie group acts then it is suggestive to use just that group as $G$. This leads us to the notion of a {\bf symmetric spectral quadruple}.

\begin{definition}
A {\bf symmetric spectral quadruple} is a spectral quadruple distinguished by
the following
conditions
\begin{enumerate}
\item $G$ is a finite dimensional Lie group
and $iH$ is then in the Lie algebra of $G$.
\item For any algebra ${A}_{{t}_{0}}$, the subgroup $K$ preserving the algebra
coincides with a maximal compact subgroup of $G$.
\item The operators $E$, $\gamma$ commute with the group $K$:
\begin{align}
[k,E]&= 0,\\
[k,\gamma]&= 0,
\end{align}
for any $k\in K$.
\end{enumerate}
\end{definition}

\begin{remark}\label{symtimechoices}
The generator $iH$ for an algebra ${A}_{0}$ is not
to be chosen as a compact generator in the Lie algebra of $G$ preserving ${A}_{0}$ as it may not fulfill the geometry-of-space requirement in Definition \ref{defspectquad} of a spectral quadruple.
\end{remark}

The examples of 1+1 dimensional de Sitter space and of finite spectral quadruples are worked out in \cite{Kopf-Paschke2000b}.

\section*{Acknowledgements}
The authors would like to thank Jos\'e M. Gracia-Bond\'{\i}a and
Florian Scheck for their encouragement and for a number of useful
discussions. T.K. most thankfully acknowledges support from the Alexander von Humboldt
Foundation.




\end{document}